# Modeling Bike Availability in a Bike-Sharing System Using Machine Learning

Huthaifa I. Ashqar[1,3], Mohammed Elhenawy[2,3], Mohammed H. Almannaa[1,3], and Ahmed Ghanem[2,3]
[1] Dept. of Civil and Environmental Engineering
[2] Dept. of Electrical and Computer Engineering
[3] Virginia Tech Transportation Institute
Blacksburg, VA 24061
{hiashqar, mohame1, almannaa, aghanem}@vt.edu

Hesham A. Rakha
Dept. of Civil and Environmental Engineering
Virginia Tech Transportation Institute
Blacksburg, VA 24061
hrakha@vt.edu

Leanna House
Dept. of Statistics
Virginia Tech
Blacksburg, VA 24061
lhouse@vt.edu

*Abstract*—This paper models the availability of bikes at San Francisco Bay Area Bike Share stations using machine learning algorithms. Random Forest (RF) and Least-Squares Boosting (LSBoost) were used as univariate regression algorithms, and Partial Least-Squares Regression (PLSR) was applied as a multivariate regression algorithm. The univariate models were used to model the number of available bikes at each station. PLSR was applied to reduce the number of required prediction models and reflect the spatial correlation between stations in the network. Results clearly show that univariate models have lower error predictions than the multivariate model. However, the multivariate model's results are reasonable for networks with a relatively large number of spatially correlated stations. Results also show that station neighbors and the prediction horizon time are significant predictors. The most effective prediction horizon time that produced the least prediction error was 15 minutes.

*Keywords—Bike prediction; bike-sharing systems; urban computing*

## I. INTRODUCTION

A growing population, with more people living in cities, has led to increased pollution, noise, congestion, and greenhouse gas emissions. One possible approach to mitigating these problems is encouraging the use of bike-sharing systems (BSSs). BSSs are an important part of urban mobility in many cities and are sustainable and environmentally friendly. As urban density and its related problems increase, it is likely that more BSSs will exist in the future due to relatively low capital and operational costs, ease of installation, pedal assistance for people who are physically unable to pedal for long distances or on difficult terrain, and better tracking of bikes [1].

One of the first BSSs in the United States was established in 1964 in Portland, with 60 bicycles available for public use. Although BSSs are still relatively limited, at present many cities, such as San Francisco and New York, have launched programs. These programs implement different payment structures, conditions, and logistical strategies. Of primary interest to this paper are those that rely on information technology (IT). One of the largest IT-based systems based in Montreal, Canada, is BIXI (BIcycle-TaXI), which employs the concept of using a bicycle like a taxi. BIXI, with its use of advanced technologies for implementation and management, illustrates a shift into the fourth generation of BSSs [2].

In 2013, San Francisco launched the Bay Area Bike Share BSS, a membership-based system providing 24-hours-per-day, 7-days-per-week self-service access to short-term rental bicycles. Members can check out a bicycle from a network of automated stations, ride to the station nearest their destination, and leave the bicycle safely locked for someone else to use [3]. The Bay Area Bike Share is designed for short, quick trips, and as a result, additional fees apply for trips longer than 30 minutes. In this system, 70 bike stations connect users to transit, businesses, and other destinations in four areas: downtown San Francisco, Palo Alto, Mountain View, and downtown San Jose [3]. Bay Area Bike Share is available to everyone 18 years and older with a credit or debit card. The system is designed to be used by commuters and tourists alike, whether they are trying to get across town at rush hour, traveling to and from Bay Area Rapid Transit (BART) and Caltrain stations, or pursuing daily activities [3].

This paper proposes an approach to modeling the number of available bikes at a bike share station using machine learning. Since the number of available bikes at a station, which has a finite number of docks, fluctuates, a repositioning (or redistribution) operation must be performed periodically. Coordinating such a large operation is complicated, time-consuming, polluting, and expensive [1]. Predicting the number of available bikes in each station over time is one of the key tasks to making this operation more efficient. In this study, Random Forest (RF) and Least-Squares Boosting (LSBoost) algorithms were used to build univariate prediction models for available bikes at each Bay Area Bike Share station. However, to reduce the number of required prediction models for the entire BSS network, we also used Partial Least-Squares Regression (PLSR) as a multivariate regression algorithm.

Following the introduction, this paper is organized into five sections. Section II briefly discusses related work from the literature, focusing on the methods proposed in previous studies. Next, a background of the regression models used is presented in Section III. In Section IV, the different data sets used in this study are described. The details of the data analysis used to construct predictive models of the number of available bikes are provided in Section V. Finally, the paper concludes with a summary of new insights and recommendations for future research on modeling the number of available bikes.

## II. RELATED WORK

The modeling of bike sharing data is an area of significant research interest. Proposed models have relied on various features, including time, weather, the built environment, and transportation infrastructure. In general, the main goals of these models have been to boost the redistribution operation [4-6], to gain new insights into and correlations between bike demand and other factors [7-10], and to support policy makers and managers in making optimized decisions [7, 11].

Froehlich, Neumann, and Oliver used four predictive models to predict the number of available bikes at each station: last value, historical mean, historical trend, and Bayesian network [12]. Two methods for time series analysis, Autoregressive Moving Average (ARMA) and Autoregressive Integrated Moving Average (ARIMA), have also been used to predict the number of available bikes/docks for each bike station. Kaltenbrunner, Meza, Grivolla, Codina, and Banchs adopted ARMA [13]; Yoon, Pinelli, and Calabrese proposed a modified ARIMA model considering spatial interaction and temporal factors [14]. However, Gallop, Tse, and Zhao used continuous and year-round hourly bicycle counts and weather data to model bicycle traffic in Vancouver, Canada [15]. That study used seasonal autoregressive integrated moving average analysis to account for the complex serial correlation patterns in the error terms and tested the model against actual bicycle traffic counts. The results demonstrated that the weather had a significant and important impact on bike usage. The authors found that the weather data (namely temperature, rain, humidity, and clearness) were generally significant; temperature and rain, specifically, had an important effect.

A multivariate linear regression analysis was used by Rixey to study station-level BSS ridership [10]. That study investigated the correlation between BSS ridership and the following factors: population density; retail job density; bike, walk, and transit commuters; median income; education; presence of bikeways; nonwhite population (negative association); days of precipitation (negative association); and proximity to a network of other BSS stations. The author found that demographics, the built environment, and access to a comprehensive network of stations were critical factors in supporting ridership.

This paper makes two major contributions to the literature. First, the univariate response models that have been used previously to predict the number of available bikes at each station ignore the correlation between stations and might become hard to implement when applied to relatively large networks. Thus, this paper investigates the use of multivariate response models to predict the number of available bikes in the network. Second, station neighbors, which are determined by a trip's adjacency matrix, are considered as significant predictors in the regression models.

## III. METHODS

In this section, we will briefly describe the three machine learning algorithms used in this paper: RF, LSBoost, and PLSR.

### A. Random Forest (RF)

Breiman proposed RF as a new classification and regression technique in supervised learning [16]. RF creates an ensemble of decision trees and randomly selects a subset of features to grow each tree. While the tree is being grown, the data are divided by employing a criterion in several steps or nodes. The correlation between any two trees and the strength of each individual tree in the forest affect, also known as the forest error rate in classifying each tree. Practically, the mean squared error of the responses is used for regression.

RF offers several advantages [16, 17]. For example, there are very few assumptions attached to its theory; it is considered to be robust against overfitting; it runs efficiently and relatively quickly with a large amount of data and many input variables without the need to create extra dummy variables; it can handle highly nonlinear variables and categorical interactions; and it ranks each variable's individual contributions in the model. However, RF also has a few limitations. For instance, the observations must be independent, which is assumed in our case.

### B. Least-Squares Boosting (LSBoost)

LSBoost is a gradient boosting of regression trees that produces highly robust and interpretable procedures for regression. LSBoost was proposed by Friedman as a gradient-based boosting strategy [18], using square loss $L(y, F) = (y - F)^2/2$, where $F$ is the actual training and $y$ is the current cumulative output $y_i = \beta_0 + \sum_{j=1}^{i-1} \beta_j h_j + \beta_i h_i = y_{i-1} + \beta_i h_i$. The new added training $\hat{F}$ is set to minimize the loss, in which the training error is computed as in [19]:

$$E = \sum_{t=1}^{N} [\beta_i h_i^t - \hat{F}^t] \qquad (1)$$

where $\hat{F}$ is the current residual error and the combination coefficients $\beta_i$ are determined by solving $\partial E / \partial \beta_i = 0$.

In this paper, RF and LSBoost were used as univariate regression techniques to model the number of available bikes in each station at any time $t$. RF and LSBoost are ensemble learning algorithms, which integrate multiple decision trees to produce robust models. However, the main difference between these two algorithms is the order in which each component tree is trained. Using randomness, RF trains each tree independently, whereas LSBoost trains one tree at a time and each new added tree is set to correct errors made by previously trained trees. The ensemble model is produced by synthesizing results from the individual trees.

### C. Partial Least-Squares Regression (PLSR)

PLSR was recently developed as a multivariate regression algorithm [20-24]. PLSR finds a linear regression model by projecting the predicted variables $Y$ and the observable variables $X$ to a new space. The basic model in the PLSR method consists of a regression between two blocks, i.e. $X$ and $Y$. Furthermore, this model contains outer relations for each of the $X$ and $Y$ blocks, and an inner relation that links both blocks. PLSR has several advantages. For example, it is suitable when the matrix of predictors $Y$ has more variables than observations, and when there is multicollinearity among observable variable $X$ values. Moreover, the PLSR method outperforms multiple linear regressions because implementing PLSR develops stable predictors. In this paper, PLSR was used as multivariate regression to reduce the number of required prediction models for the number of available bikes at any time $t$ for the entire BSS network.

## IV. DATA SET

This study used anonymized bike trip data collected from August 2013 to August 2015 in the San Francisco Bay Area as shown in Fig. 1 [25]. This study used two data sets. The first data set includes station ID, number of bikes available, number of docks available, and time of recording. The time data include year, month, day of the month, time of day, and minutes at which a record was documented. As an incident was documented every minute for 70 stations in San Francisco over 2 years, this data set contains a large number of recorded incidents. This data set was exposed to a change-detection process to determine times when a change in bike count occurred at each station. From this data set, as a result of pre-processing, the station ID, number of bikes available, month, day of the week, and time of day were extracted for use as features. Subsequently, each station's ZIP code was assigned and input to the set.

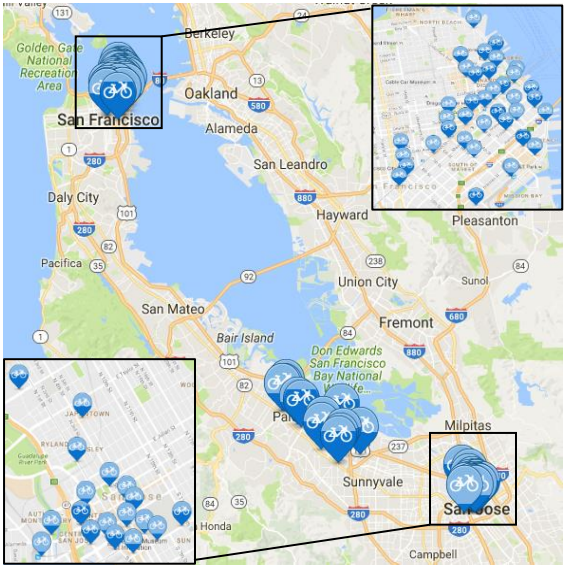

Fig. 1. Stations map.

The second data set contains different attributes: the date (in month/day/year format), ZIP code, and 22 other variables describing the daily weather for each ZIP code over the 2-year period. The number of available bikes at station $i$ at time $t$, the number of available bikes at its neighbors at the same time $t$, month of the year, day of the week, and time of day were all extracted from the two data sets as parameters that affect the model. Specifically, the neighbors of a station $i$ were defined based on the number of trips originated from station $j$, in which $j \neq i$, and ended at station $i$. In that sense, we generated the adjacency matrix of the BSS network and found the highest 10 in-degree stations for station $i$, which were assigned as neighbors of station $i$. In addition, an unpublished work by the authors [26] investigated various weather data as predictors to determine the reasonable parameters that mainly affect the prediction models. From the weather information, mean temperature, mean humidity, mean visibility, mean wind speed, precipitation, and events in a day (i.e., rainy, foggy, or sunny) were all selected. These parameters were selected based on subject-matter expertise and previous related studies [9, 15], and they were found to be significant in predicting the number of available bikes at Bay Area Bike Share stations [26].

## V. DATA ANAYSIS AND RESULTS

### A. Univariate Models

RF and LSBoost algorithms were applied to create univariate models to predict the number of available bikes at each of the 70 stations of the Bay Area Bike Share network. The two algorithms were applied to investigate the effect of several variables on the prediction of the number of available bikes in each station $i$ in the Bay Area BSS network, including the available bikes at station $i$ at time $t$, the available bikes at its neighbors at the same time $t$, the month of the year, day of the week, time of day, and various selected weather conditions. The predictors' vector for station $i$ at time $t$, denoted by $X_t^i$, was used in the built models to predict the $log$ of the number of available bikes at station $i$ at time $t$ and at a prediction horizon time, denoted by $\log(y_{t+\Delta}^i)$, where $i = 1, 2, \ldots, 70$. The effect of different prediction horizons, $\Delta$ (range 15–120 minutes), on the performance of both algorithms was investigated by finding the Mean Absolute Error (MAE) per station (i.e., bikes/station), which can be described as the prediction error. Moreover, as the number of generated trees by RF and LSBoost is an important parameter in implementing both algorithms, we investigated its effect by changing the number of generated trees from 20 trees to 180 trees with a 40-tree step.

As shown in Fig. 2 and Fig. 3, the prediction errors of RF and LSBoost increase as the prediction horizon $\Delta$ increases. The lowest prediction error for both algorithms occurred at a 15-minute prediction horizon. Moreover, the prediction error of RF and LSBoost decreases as the number of trees increases until it reaches a point where increasing the number of trees will not significantly improve the prediction accuracy. Fig. 2 and Fig. 3 also show that a model consisting of 140 trees yields a relatively sufficient accuracy.

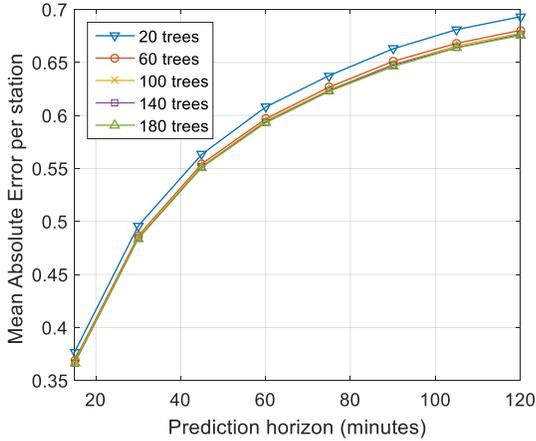

Fig. 2. RF MAE at different prediction horizons and number of trees.

Comparing the two algorithms, the models produced by RF generally have a smaller prediction error than those produced by LSBoost. LSBoost is a gradient-boosting algorithm, which usually requires various regularization techniques to avoid overfitting [27]. As Fig. 3 clearly shows, as the prediction horizon time increases, the prediction error increases (this is also clearly shown in Fig. 5 in the next section).

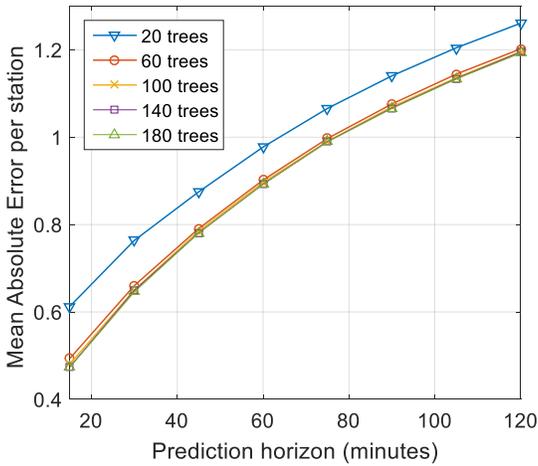

Fig. 3. LSBoost MAE at different prediction horizons and number of trees.

### B. Multivariate Models

PLSR was used as a multivariate regression to reduce the number of required prediction models for bike stations in the BSS network. When a BSS network has a relatively large number of stations, tracking all the specified models for each bike station becomes complex and time-consuming. For that reason, we examined the adjacency matrix of the Bay Area BSS network and found that the network can be divided into five regions as shown in Fig. 4. In fact, the bike stations that resulted from the adjacency matrix in each region were found to share the same ZIP code. This means that the majority of bike trips occurred within the same region and very few trips went from one region to another.

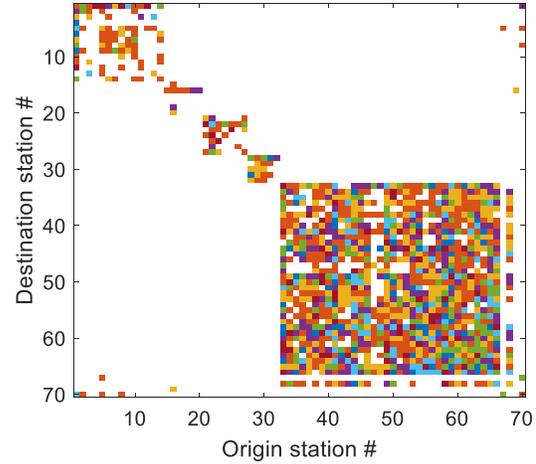

Fig. 4. Adjacency matrix of the Bay Area Bike Share network.

Using PLSR as a regression algorithm can build prediction models for multivariate response. Therefore, PLSR was applied to reduce the number of models to five, each of which is specified for one region (i.e., one ZIP code) to reflect the spatial correlation between stations. The input predictors' vector is $X_t^i$, which consists of the available bikes at the station $i$ at time $t$, the available bikes at its neighbors at the same time $t$, the month of the year, day of the week, time of day, and various selected weather conditions. The response's vector is $\log(Y_{t+\Delta}^i)$, where $i = 1, 2, 3, 4, 5$, which is the log of the number of available bikes at all stations in each of the studied regions at a prediction horizon time $\Delta$ (range 15–120 minutes). We found that the prediction errors for PLSR were higher than the RF and LSBoost prediction errors when $\Delta = 15$ minutes, as shown in Fig. 5. Although the prediction errors resulting from PLSR were higher than the previous results, the resulting models from PLSR are sufficient and desirable for relatively large BSS networks.

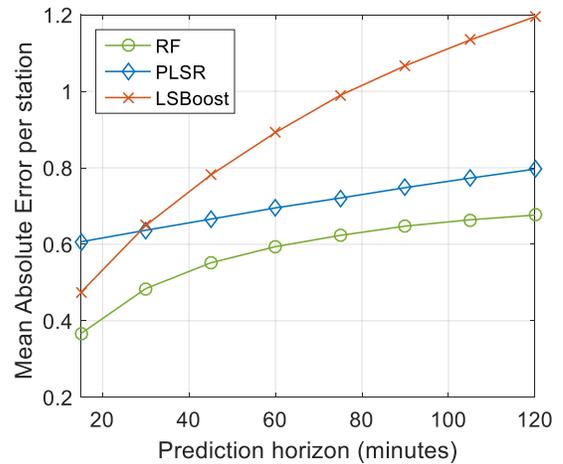

Fig. 5. PLSR, RF, and LSBoost MAE at different prediction horizons.

## VI. Conclusions and Recommendations for Future Work

In this paper, we modeled the number of available bikes at San Francisco Bay Area Bike Share stations using machine learning algorithms. The investigation applied two approaches: using univariate regression algorithms, RF and LSBoost, and using a multivariate regression algorithm, PLSR. The univariate models were used to model the available bikes at each station. RF with an MAE of 0.37 bikes/station outperformed LSBoost with an MAE of 0.58 bikes/station. On the other hand, the multivariate model, PLSR, was applied to model available bikes at the spatially correlated stations of each region obtained from the trips adjacency matrix. Results clearly show that the univariate models produced lower error predictions compared to the multivariate model, in which the MAE was approximately 0.6 bikes/station. However, the multivariate model's results might be acceptable and reasonable when modeling the number of available bikes in BSS networks with a relatively large number of stations.

Investigating BSS networks in terms of determined regions gives new insights to policy makers. The fact that stations in each region derived by the multivariate analysis share the same ZIP code implies that most of the trips were short distance, which may be influenced by the overtime fees applied when trips are longer than 30 minutes. The results also illustrate that station neighbors, prediction horizon time, and weather variables (e.g., temperature and humidity) were found to be significant in modeling the number of available bikes. Specifically, when the prediction horizon time increases, the prediction error increases, with the most effective prediction horizon being 15 minutes. Determining prediction horizon is beneficial to policy makers and technicians to learn how to manage the BSS more responsively, and achieve better performance in prediction. Future work could model the number of available bikes by adding memory as a predictor to handle information related to the number of available bikes in the past.

### Acknowledgment

This research effort was funded by the UrbComp program from the National Science Foundation.